# Ferroelectricity Driven by Twisting of Silicate Tetrahedral Chains

Hiroki Taniguchi[1*], Akihide Kuwabara[2], Jungeun Kim[3], Younghun Kim[4], Hiroki Moriwake[2], Sungwng Kim[5], Takuya Hoshiyama[6], Tsukasa Koyama[6], Shigeo Mori[6], Masaki Takata[3,4,7], Hideo Hosono[1,8], Yoshiyuki Inaguma[9], and Mitsuru Itoh[1]

[1]Materials and Structures Laboratory, Tokyo Institute of Technology, Yokohama 226-8503, Japan.

[2]Nanostructures Research Laboratory, Japan Fine Ceramics Center, Nagoya 456-8587, Japan.

[3]Japan Synchrotron Radiation Research Institute, Sayo-gun, Hyogo 679-5198, Japan.

[4]Graduate School of Frontier Sciences, The University of Tokyo, Kashiwa, Chiba 277-8561, Japan.

[5]Department of Energy Science, Sungkyunkwan University, Jangan-Gu, Suwon 440-746, Korea.

[6]Department of Materials Science, Osaka Prefecture University, Sakai, Osaka 599-8531, Japan.

[7]RIKEN Harima Institute, Sayo, Hyogo 679-5148, Japan.

[8]Frontier Research Center, Tokyo Institute of Technology, Yokohama 226-8503, Japan.

[9]Faculty of Science, Gakushuin University, Toshima-ku, Tokyo 171-8588, Japan.

*Correspondence to: taniguchi.h.aa@m.titech.ac.jp

**Conventional perovskite-type ferroelectrics are based on octahedral units of oxygen, and often comprise toxic Pb to achieve robust ferroelectricity. Here, we report the ferroelectricity in a silicate-based compound, $Bi_2SiO_5$ (BSO), induced by a structural instability of the corresponding silicate tetrahedral chains. A low-energy phonon mode condenses at ~ 673 K to induce the proper ferroelectric phase transition. Polarization switching was observed in a BSO single crystal with a coercive field of 30 kV/cm and a spontaneous polarization of 0.3 $\mu C/cm^2$ along a direction normal to the cleavage plane. The in-plane polarization was estimated by first principles calculations to be 23 $\mu C/cm^2$. The present findings provide a new guideline for designing ferroelectric materials based on $SiO_4$ tetrahedral units, which is ubiquitously found in natural minerals.**

Because of its diverse functionality, ferroelectricity is a key ingredient for electronic and optical technologies applied to non-volatile memory, actuators and sensors. (*1*) Ferroelectric materials are characterized by a switchable polarization caused by spontaneous displacements of cations relative to anions, which generally appears through the ferroelectric phase transition from centrosymmetric to non-centrosymmetric phases. In the case of so-called displacive-type phase transitions, the phase transition is driven by the freezing of a zone-centre optical phonon mode, "soft mode", using an eigenvector similar to the polar displacement observed in the ferroelectric phase. (*2-4*) Recent studies have clarified that the soft mode instability in conventional perovskite-type oxides is caused by covalency between the cations and oxygen ions in the octahedral oxygen units. (*5-7*) Current ferroelectric applications thus rely heavily upon toxic Pb-based compounds, because the large covalency of Pb-O is favourable for causing structural instability in these compounds to induce robust ferroelectricity. Alternately, the unconventional ferroelectricity observed in Bi-based perovskite-type oxides like $BiFeO_3$ (BFO), has recently attracted attention in terms of the high-$T_c$ and the large spontaneous polarization achieved without Pb. The mechanism has been understood as a lone-pair ordering, where the 6*s* orbital of Bi hybridizes with the 2*p*-orbital of an adjacent oxygen to induce the large polar displacement of Bi. (*8,9*) Also in this case, the ferroelectricity stems from the unique property of the specific element, Bi. Such a strong element-dependence of current ferroelectric oxides provides a demand for a new guideline principle for designing ferroelectric materials. (*10,11*)

Here, we report for the first time the ferroelectric phase transition driven by the structural instability of the pyroxene-type one-dimensional chains of oxygen tetrahedra, which is found in the silicate-based compound $Bi_2SiO_5$ (BSO). BSO has a layered structure, which comprises a single layer of one-dimensional silicate chains sandwiched between $Bi_2O_2$ sheets (Fig. 1a). The proper ferroelectric phase transition at 673 K is induced by the freezing of the soft mode that stems from twisting of the silicate chains. The ferroelectricity is verified by a direct observation of *P-E* hysteresis loops. This mechanism arises from the structural instability of the tetrahedral chain abundant in the earth's crust. Therefore, the present finding thus offers a new route for designing environmentally friendly ferroelectric oxides that are free from the restriction of specific elements.

The crystal structure of BSO has previously been reported to be an orthorhombic $Cmc2_1$ structure at room temperature. (*12*) Our transmission electron microscopy (TEM) observations, however, revealed that the crystallographic *β* angle of BSO slightly deviates from 90° at room temperature, indicating a monoclinic distortion of BSO (Supplementary Fig. 1a). Structural analyses employing X-ray powder diffraction indicate that BSO has a monoclinic *Cc* crystal structure with lattice parameters of $a$ = 15.1193(1) Å, $b$ = 5.4435(1) Å, $c$ = 5.2892(1) Å, and $β$ = 90.0695(20)° (Supplementary Fig. 2a). At 793 K, however, the *β* angle is found to change to 90°, demonstrating the structural phase transition of BSO from the monoclinic phase to an orthorhombic phase with elevating temperature (Supplementary Fig. 1b). The high-temperature orthorhombic phase was also investigated by X-ray powder diffraction at 773 K. Based on Rietveld analyses, it was established that BSO has an orthorhombic *Cmcm* structure with lattice parameters of $a$ = 15.1968(1) Å, $b$ = 5.4843(1) Å, and $c$ = 5.2964(1) Å (Supplementary Fig. 2b). In the high-temperature phase of BSO, as shown in Fig. 1b, the ⋯O1-O1-O1⋯ bonds that form the one-dimensional chains of the $SiO_4$ tetrahedral units are linear. As the temperature decreases, the bonds twist with an angle of 172.2°, which is accompanied by a coherent displacement of the Bi ions, and a distortion of the O3 grid patterns. The spontaneous polarization along the *c*-direction in the low-temperature phase is consequently induced by the relative displacement of the anionic $SiO_4$ tetrahedral units and cationic $Bi_2O_2$ cationic layers. Synchronizing with them, the Bi ions slightly shift to a direction normal to the O3 grids, thereby inducing the polarization along the *a*-direction.

Figure 2a presents the temperature dependence of the dielectric permittivity in a BSO ceramic measured at 1 kHz. The dielectric permittivity gradually increases with elevating temperature and culminates at 673 K, indicating the onset of the ferroelectric phase transition. A subsequent cooling process shows a small hysteresis in the peak anomaly, as denoted by the dotted lines in the figure, manifesting in a phase transition that is of a weak first-order. A slight discrepancy in the dielectric permittivity between the heating and the cooling would be due to an ionic conduction, which was suggested by an increasing tan δ observed in the high-temperature region. The polarization switching is demonstrated in the single crystal of BSO along the *a*-direction that is normal to the cleavage plane of the single crystal plate, where an applied voltage was swept from ± 35 to ± 65 kV/cm. As seen in Fig. 2b, a well-defined $P$-$E$ hysteresis loop was observed at 573 K, confirming the occurrence of ferroelectricity in BSO. Note that an accurate estimation of the saturated polarization and the coercive field was not available because of the sample damage at the large bias field. Nevertheless, the present result indicates that the spontaneous polarization projected onto the $a$-direction, $P_s^a$, is not so much larger than 0.3 μC/cm$^2$. The small out-of-plane polarization agrees well with the slight monoclinic distortion of BSO at the ferroelectric phase.

Raman scattering experiments were also performed on the BSO single crystals to elucidate the origin of the ferroelectricity in BSO. Fig. 3a shows low-frequency Raman spectra of BSO observed with increasing temperature from 83.15 K to 723.15 K. As indicated by the red arrow in the figure, the low-frequency peak observed around 70 cm$^{-1}$ at 83.15 K, was found to decrease in frequency to zero with elevating temperature. The peak completely vanishes on further heating, confirming the polar character of this mode since the polar mode is Raman-inactive in the centrosymmetric $Cmcm$ phase. This is the ferroelectric soft mode in BSO. In the proper displacive-type ferroelectric phase transition, it is known that the soft mode softens by obeying Cochran's law, $\omega_{\text{soft}} = C\,|T - \tilde{T}_c|^{0.5} + \Delta\omega$, where $\omega_{\text{soft}}$, $C$, $T_c$, and $\Delta\omega$ are the soft mode frequency, the Curie constant, the phase transition temperature, and the frequency offset for the first-order phase transition, respectively. (*2,3*) Figure 3b shows a temperature dependence of $\omega_{\text{soft}}$ examined by fitting of the spectra, where data points close to 673 K were omitted because of an uncertainty in the analyses due to overdamping of the soft mode. To obtain the exact $T_c$ that is estimated from the Raman spectra, we examined the temperature dependence of the peak at around 300 cm$^{-1}$, which is indicated by the blue arrow in Fig. 3a. The temperature dependence of the peak position presented in Fig. 3c shows a distinct cusp at 673 K, corresponding to $T_c$. The $T_c$ value agrees with the anomaly in the dielectric permittivity shown in Fig. 2a. The temperature dependence of the soft mode frequency was then examined using Cochran's law with a fixed $T_c$ = 673 K, leading to a good fit with the parameters of $C$ = 2.4 and $\Delta\omega$ = 12.1 cm$^{-1}$, as indicated by a solid line in Fig. 3b. This result confirms the displacive-type proper ferroelectric phase transition of BSO, which is driven by the ferroelectric soft mode.

Phonon dispersion curves in the high-temperature $Cmcm$ phase obtained by first principles calculations are shown in the left panel of Fig. 4a. In the first principles calculation, the ferroelectric soft mode is characterized by the soft mode having an imaginary frequency at the Γ-point in the paraelectric phase due to instability of its vibration in the centrosymmetric structure. (*14*) As seen in the figure, the calculation clarifies the existence of the soft mode in the $Cmcm$ phase, as indicated by the arrows. A discontinuity in the soft mode dispersion toward the Z-point is caused by the splitting of the longitudinal and transverse components, indicating that the soft mode polarizes along the crystallographic $c$-direction. In the phonon dispersion curves for the $Cc$ phase, however, the soft mode vanishes, confirming the crystal symmetry of $Cc$ for the ferroelectric ground state of BSO. The displacement pattern of the soft mode is presented in Fig. 4b with the equilibrium structure of the $Cmcm$ phase, where the corresponding animation is indicated in the Supplementary Movie S1. The equilibrium paraelectric structure in the central panel, which is obtained by first-principles calculations, is consistent with that determined by X-ray powder diffraction, as shown in Fig. 1b. The displacement pattern in the (100) plane is found to be composed of twisting silicate tetrahedral chains, which are accompanied by the deformation of the oxygen grid pattern. Synchronizing with them, the Bi ions, however, show a slight shift along the crystallographic $c$-axis. At temperatures sufficiently higher than $T_c$, the silicate tetrahedral chains alternate between opposing twist configurations. As the temperature approaches $T_c$, the frequency decreases toward zero and polarization is induced in BSO when it freezes at $T_c$. The net polarization is finally generated cooperatively with the concomitant distortions of the $Bi_2O_2$ sublattice. Note that the freezing of the calculated soft mode displacement induces the symmetry reduction from the $Cmcm$ structure to the $Cmc2_1$ structure. Furthermore, a small monoclinic distortion occurs at $T_c$, probably because of a smaller total energy of $Cc$ than that of $Cmc2_1$, where the difference is estimated from first-principles calculations to be 0.2 meV/$Bi_2SiO_5$. In addition to the Γ-point, several unstable modes are observed at the Y- and S-points. According to the calculation, the Γ-point distortion is the energetically most favourable point among all corresponding distortions for the unstable modes (Supplementary Table S2). Because of a restriction of the very thin crystal shape that allows only the polarization measurement along the $a$-axis (the inset of Fig. 2A), the value of the in-plane polarization along the $c$-axis was not successfully obtained in the present study. Instead, it was estimated from Born effective charges and the optimized $Cc$ structure using first principles calculations. The calculated value of $P_s^c \sim 23$ μC/cm$^2$ along the $c$ direction is comparable with that of the typical perovskite-type ferroelectric material, $BaTiO_3$ (~ 25 μC/cm$^2$). This result indicates the potential of the tetrahedral chain for achieving robust ferroelectricity. Note that the calculated structural parameters, Born effective charges, and permittivity tensors are indicated in the Supplementary Tables S3, S4, and S5. The ratios of lattice constants agree well with the experimental results, although the individual values are slightly underestimated by ~ 2 %.

The twisting fluctuation of the silicate-chain, which mainly composes the soft mode displacement in BSO, is also found in the framework of certain silicates, such as cristobalite, the high-temperature form of quartz. In this case, the structure is composed only of silicate chains, which are connected to each other in parallel. Interestingly, the *α-β* transition of cristobalite is driven by the twisting of the silicate chain, which is similar to the case of BSO. (*15*) This suggests that the silicate chains themselves possess the unstable twisting fluctuation to induce the structural phase transition. Thus, we believe that the ferroelectric phase transition of BSO stems intrinsically from the structural instability of the silicate chains, even though the distortion of the $Bi_2O_2$ sub-lattice also takes part in the displacement of the ferroelectric soft mode. Finally, the ferroelectric instability in BSO appears based on the oxygen-tetrahedral units. This is distinct from traditional ferroelectric materials, where the ferroelectricity is designed by the arrangement of the oxygen-octahedral-units and interstitial cations. The present result sheds light on a development of "tetrahedral-engineering" for the ferroelectrics, beyond the conventional "octahedral-engineering".

In summary, the twisting of the silicate tetrahedral chains is found to induce the switchable polarization in the silicate-based compound, BSO. The high $T_c$ and the relatively large spontaneous polarization indicate the potential for use of one-dimensional tetrahedral chains for designing ferroelectric oxides. Furthermore, these one-dimensional tetrahedral chains can be found in many ubiquitous oxides, such as silicates and aluminates, which are abundant in the earth's crust. (*16*) We believe that the present findings will help provide new guidelines for the development of environmentally friendly ferroelectric oxides.

**Acknowledgements:** This work was supported by the Ministry of Education, Culture, Sports, Science, and Technology of Japan through a Grant-in-Aid for Young Scientists (B) (No. 24760543) and Scientific Research (C) (No. 24560833), and Hitachi Metals - Materials Science Foundation.

**References**

1. G. H. Haertling, Ferroelectric ceramics: history and technology. *J. Am. Ceram. Soc.* **82**, 797–818 (1999).
2. W. Cochran, Crystal stability and the theory of ferroelectricity. *Adv. Phys.* **9**, 387 (1960).
3. W. Cochran, Crystal stability and the theory of ferroelectricity part II. Piezoelectric crystals. *Adv. Phys.* **10**, 401 (1961).
4. J. F. Scott, Soft-mode spectroscopy: Experimental studies of structural phase transitions. *Rev. Mod. Phys.* **46**, 83 (1974).
5. R. E. Cohen, Origin of ferroelectricity in perovskite oxides. *Nature* **358**, 136-138 (1992).
6. Y. Kuroiwa *et al.*, Evidence for Pb-O Covalency in Tetragonal $PbTiO_3$. *Phys. Rev. Lett.* **87**, 217601 (2001).
7. H. Taniguchi *et al.*, Mechanism for suppression of ferroelectricity in $Cd_{1-x}Ca_xTiO_3$. *Phys. Rev. B* **84**, 174106 (2011).
8. R. Seshadri, N. A. Hill, Visualizing the Role of Bi 6s "Lone Pairs" in the Off-Center Distortion in Ferromagnetic $BiMnO_3$, *Chem. Mater.* **13**, 2892-2899 (2001).
9. J. B. Neaton *et al.*, First-principles study of spontaneous polarization in multiferroic $BiFeO_3$, *Phys. Rev. B.* **71**, 014113 (2005).
10. Y. Saito *et al.*, Lead-free piezoceramics. *Nature* **432**, 84-87 (2004).
11. P. K. Panda, Review: environmental friendly lead-free piezoelectric materials. *J. Mater. Sci.* **44**, 5049-5062 (2009).
12. J. Ketterer, V. Krämer, Crystal structure of the bismuth silicate $Bi_2SiO_5$. *N. Jb. Miner. Mh.* **H. 1**, 13-18 (1986).
13. H. Taniguchi *et al.*, Critical soft-mode dynamics and unusual anticrossing in $CdTiO_3$ studied by Raman scattering. *Phys. Rev. B* **76**, 212103 (2007).
14. R. E. Cohen, H. Krakauer, Lattice dynamics and origin of ferroelectricity in $BaTiO_3$: Linearized-augmented-plane-wave total-energy calculations. *Phys. Rev. B* **42**, 6416–6423 (1990).
15. Dorian M. Hatch and Subrata Ghose, The α-β transition in cristobalite, $SiO_2$, *Phys. Chem. Minerals* **17**, 554 (1991).
16. F. Liebau, Structural chemistry of silicates: structure, bonding, and classification 1-6 (Springer-Verlag, 1985).

**Figure Captions:**

**Fig. 1**. (**A**) The crystal structure of $Bi_2SiO_5$ (BSO) in the ferroelectric phase, which is obtained from the structural analyses. The one-dimensional chains of the $SiO_4$ tetrahedra running along the *c*-axis are sandwiched by $Bi_2O_2$ layers. The cleavage plane is normal to the *a*-axis. (**B**) The crystal structures of BSO projected onto the *b-c* plane. The upper and lower figures show the high-temperature (*Cmcm*) and low-temperature (*Cc*) structures, respectively. The ···O1-O1-O1··· chain is twisted by 172.2° in the *Cc* phase, which is in contrast to the linear configuration shown in the *Cmcm* phase.

**Fig. 2.** (**A**) The temperature dependence of the dielectric permittivity and tan δ of BSO measured at 1 kHz. The red and blue arrows in the figure denote the heating and subsequent cooling processes, respectively. The culminated temperatures for the heating and cooling processes are indicated by broken lines. (**B**) The *P-E* hysteresis loop of BSO at 300 K was measured using a Sawyer-Tower circuit. The applied voltage was swept from ± 35 to ± 65 kV/cm.

**Fig. 3.** (**A**) The temperature dependence of the Raman spectrum of BSO observed with elevating temperature from 83.15 to 723.15 K. The inset shows the single crystal plate of BSO used in the present study. The scattering configuration for the Raman experiment is –*a*(*cc*)*a* in the Porto's notation, where the crystallographic axes are indicated in the inset. The ferroelectric soft mode in BSO, which is denoted by the red arrow in the low-frequency region, shows a decrease in frequency toward zero with elevating temperature. The peak around 300 cm$^{-1}$, as indicated by the blue arrow, is used to determine the precise transition temperature (see panel **C**). (**B**) The temperature dependence of the soft mode frequency was analysed by a spectral fitting using the damped harmonic oscillator (DHO) model, $I(\omega) = A \cdot (\omega_0^2 \Gamma \omega / ((\omega_0^2 - \omega^2)^2 + \Gamma^2 \omega^2))$, where *A*, $\omega_0$, and Γ denote the amplitude, the harmonic frequency, and the damping factor of the oscillation, respectively. The soft mode frequency $\omega_s$ is calculated by $(\omega_0^2 - \Gamma^2)^{1/2}$ according to Ref. (*13*). The curve shown in panel **B** is calculated using Cochran's law (see text for details). (**C**) The temperature dependence of the frequency for the peak around 300 cm$^{-1}$ was obtained by a spectral fitting using the DHO model. The cusp at 673 K indicates the ferroelectric phase transition temperature of BSO.

**Fig. 4.** (**A**) Phonon dispersion curves of BSO at the paraelectric *Cmcm* (left) and *Cc* (right) phases were obtained by first principles calculations. Arrows indicate the ferroelectric soft mode at the Γ-point of the Brillouin zone. (**B**) The displacement pattern for the ferroelectric soft mode. The central illustration presents the equilibrium structure of BSO at the *Cmcm* phase, which is determined by first principles calculations. The illustrations are projected onto the *b-c* plane with selected ions for simplicity.

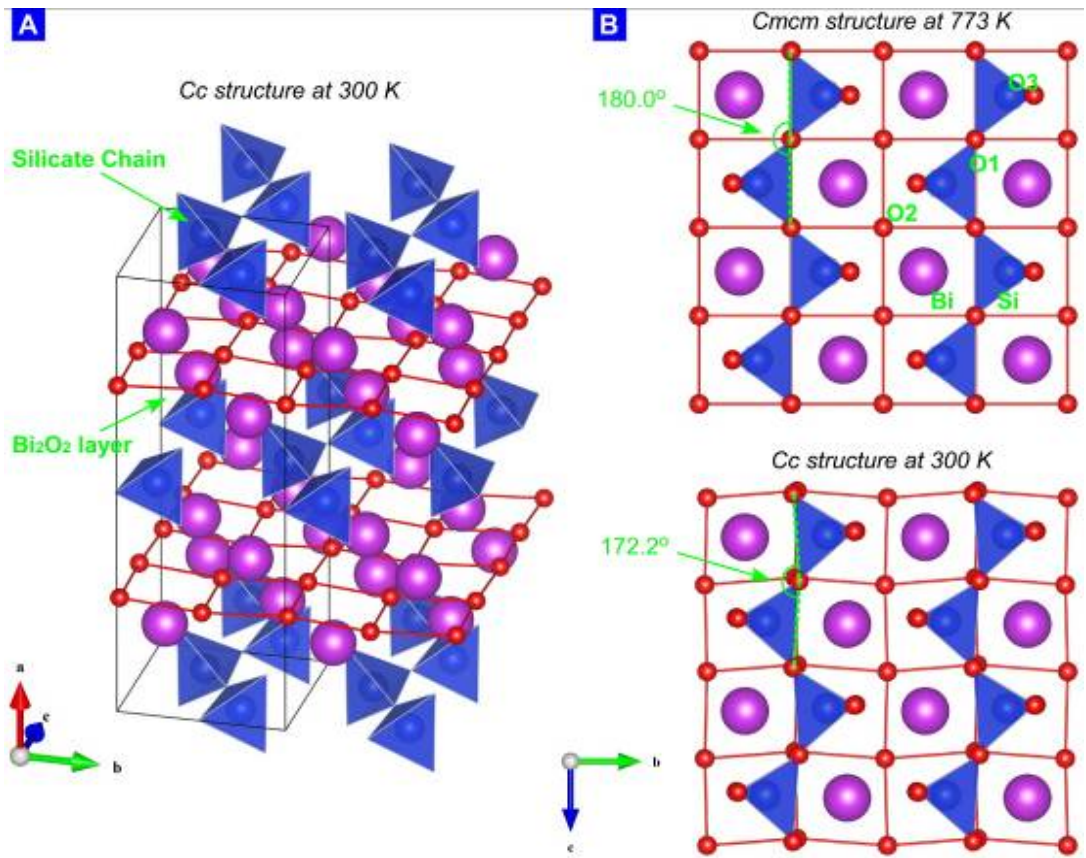

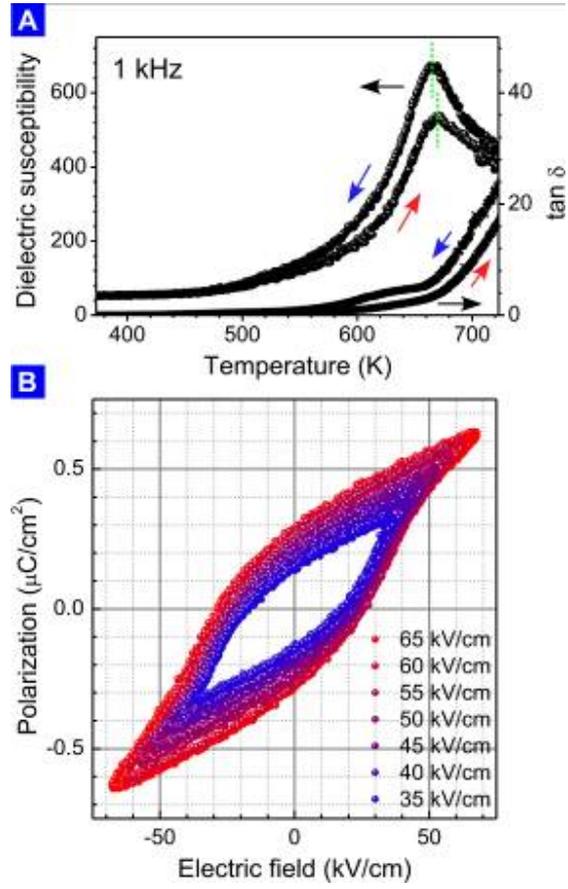

H. Taniguchi *et al.* Figure 2

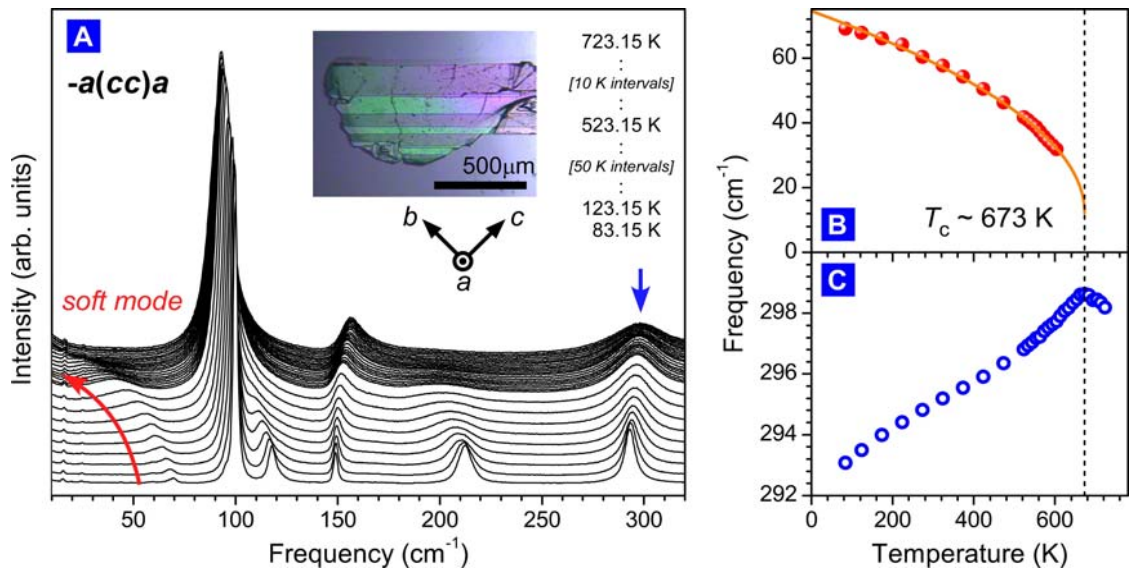

H. Taniguchi *et al.* Figure 3

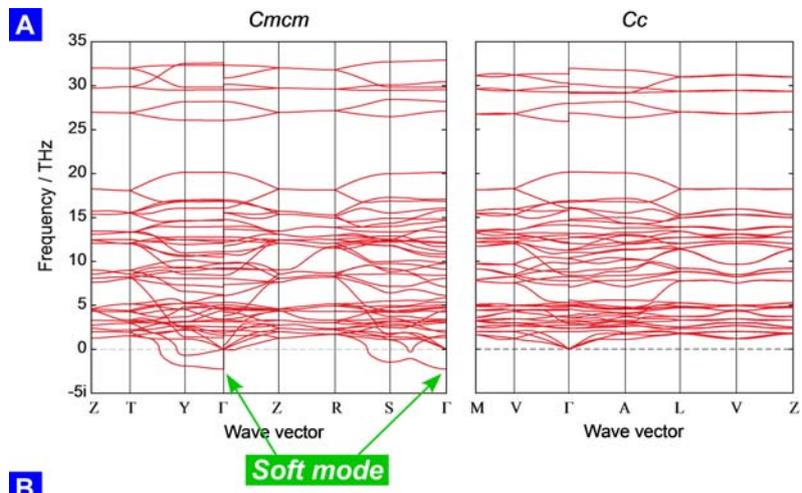
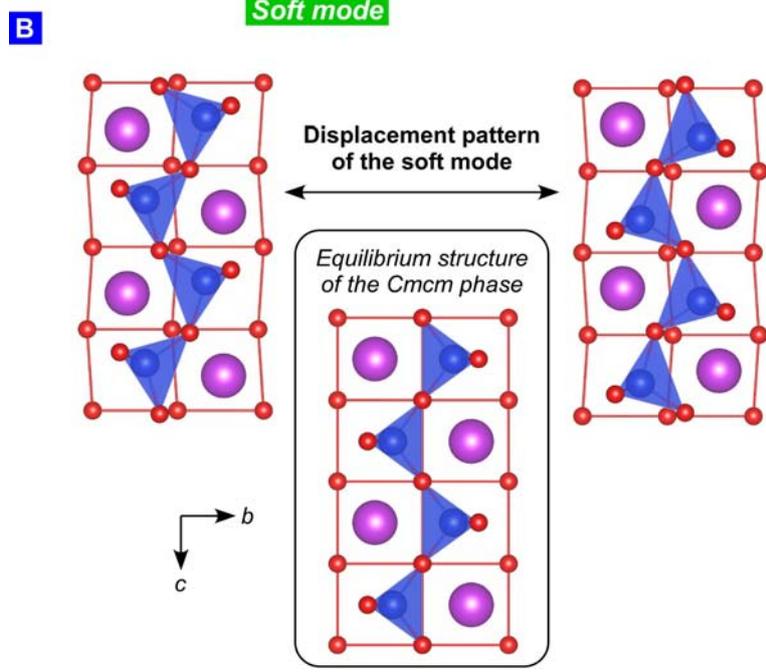

H. Taniguchi *et al.* Figure 4

**Supplementary Materials:**

**Materials and Methods:**

In the present study, we synthesized BSO single crystals by crystallization from a melt: $Bi_2O_3$ and $SiO_2$ powders with a 1:1 molar ratio were mixed together in an agate mortar, and heated in a Pt crucible at 1373K over a period of 2h. The resulting melt was subsequently cooled at a rate of 10 K/h to 1073K, followed by furnace cooling until room temperature was reached. Finally, we obtained BSO thin plates with (100) cleavage planes (Fig. 1b). As seen in the figure, twin structures of several hundred microns are clearly observed in the polarized microscope images. We also fabricated BSO ceramics by Spark plasma Sintering (SPS) for the BSO powder, which was prepared by grinding the single crystals.

The present micro-Raman scattering experiments were performed with an $Ar^+$ laser of wavelength, 514.5 nm, at ~ 10mW, which was focused onto the BSO single crystal by a 50× objective lens (N. A. = 0.55) to select a single domain area. The scattered light was collected with a back scattering geometry and analysed using a polarizer, where a scattering configuration was -$a(cc)a$ according to Porto's notation (see Fig. 3a). Raman spectra were obtained using a Horiba-Jovin-Yvon T64000 Raman spectrometer equipped with a liquid $N_2$ cooled charge-coupled device (CCD) camera. The temperature of the sample was controlled using a Linkam Scientific Instruments THMS600 microscope stage with an accuracy of ±0.1 K.

First-principles calculations were performed using the projector augmented wave (*1*) method and the local density approximation (LDA) (*2*) as implemented in the VASP code (*3*). Plane waves up to 500 eV were used as basis sets. Valence electron configurations of the potentials were taken as $5d^{10} 6s^2 6p^3$ for Bi, $3s^2 3p^2$ for Si, and $2s^2 2p^4$ for O. The radial cutoffs of the PAW potentials of Bi, Si and O were 2.50, 1.60 and 1.52 Å, respectively.

Primitive cells of $Bi_2SiO_5$ high-temperature (*Cmcm*) and low-temperature (*Cc*) phases were optimized with 5 × 5 × 5 *k*-point sampling meshes based on the Monkhorst-Pack scheme (*4*). This provided an energy convergence of less than 1 meV per $Bi_2SiO_5$ formula unit (f.u.). Crystal structures were fully relaxed until residual forces were less than 0.0002 eV/Å. Phonon calculations were performed using the Parlinski-Li-Kawazoe method (*5*) as implemented in the PHONOPY code (*6*). Interatomic force constants in real space were obtained from 4 × 4 × 1 supercells of the conventional unit cells of the *Cmcm* and *Cc* phases. In the calculations of the supercells, a *k*-point at (1/4 1/4 1/4) was chosen for sampling in the Brillouin zone. Non-equivalent atomic displacements of 0.01 Å were employed for calculations of the force constants based on symmetry analyses. A non-analytical term correction (*7,8*) was added to dynamical matrices to include an effect of LO-TO splitting near the Γ point. Born effective charges and dielectric constants were obtained using the linear response approach (*9*).

X-ray powder diffraction measurements on the powder samples of $Bi_2SiO_5$ were carried out at the powder diffraction beamline BL02B2 (Proposal No. 2011B2089) of SPring-8 to obtain the pattern of high statistics with high angular resolution ($d$ < 0.5 Å). To obtain a homogeneous intensity distribution in the Debye-Scherrer powder ring, the as-grown powders were carefully crushed into fine powders and sealed in 0.1 mm$\phi$ glass-capillaries. The temperature was controlled using a $N_2$ gas flowing system in the temperature range from 300–1000 K. The wavelength of the incident X-rays was 0.35206(1) Å, which was calibrated with the use of standard $CeO_2$ powders obtained from NIST. The diffraction patterns were collected for 55 minutes on an imaging plate installed on a large Debye-Scherrer camera. (*10*) The structural parameters of the room temperature ferroelectric phase and the high temperature phase $Bi_2SiO_5$ are summarized in the Supplementary Table 1.

Transmission electron microscopy (TEM) experiments were carried out using a JEM-2010 transmission electron microscope with an accelerating voltage of 200 kV in the temperature range between 298 K and 800 K. We took various electron diffraction (ED) patterns at 793 K and 298 K and carefully examined the changes of the ED patterns to clarify the crystal structures in the high-temperature and low-temperature phases. Note that the reflection spots in the ED patterns (Supplementary Fig. 1) are indexed on the basis of the high-temperature (*Cmcm*) structure.

All crystal structures were produced using the visualization program VESTA. (*11*)

References:

1. G. Kresse, D. Joubert, From ultrasoft pseudopotentials to the projector augmented-wave method. *Phys. Rev. B* **59**, 1758-1775 (1998).
2. D. M. Ceperley and B. J. Alder, Ground state of the electron gas by a stochastic method. *Phys. Rev. Lett.* **45**, 566-569 (1980).
3. G. Kresse, J. Furthmüller, Efficient iterative schemes for *ab initio* total-energy calculations using a plane-wave basis set. *Phys. Rev. B* **54**, 11169-11186 (1996).
4. H. J. Monkhorst, J. D. Pack, Special points for Brillouin-zone integrations. *Phys. Rev. B* **13**, 5188 (1976).
5. K. Parlinski *et al.*, First-Principles Determination of the Soft Mode in Cubic $ZrO_2$. *Phys. Rev. Lett.* **78**, 4063 (1997).
6. A. Togo *et al.*, First-principles calculations of the ferroelastic transition between rutile-type and $CaCl_2$-type $SiO_2$ at high pressures. *Phys. Rev. B* **78**, 134106 (2008).
7. X. Gonze, C. Lee, Dynamical matrices, Born effective charges, dielectric permittivity tensors, and interatomic force constants from density-functional perturbation theory. *Phys. Rev. B* 55, 10355 (1997).
8. Y. Wang *et al.*, A mixed-space approach to first-principles calculations of phonon frequencies for polar materials. *J. Phys.: Condens. Matter.* **22**, 202201 (2010).
9. M. Gajdoš *et al.*, Linear optical properties in the projector-augmented wave methodology. *Phys. Rev. B* **73**, 045112 (2006).
10. E. Nishibori *et al.*, The large Debye–Scherrer camera installed at SPring-8 BL02B2 for charge density studies, *Nucl. Instr. Meth. A* **467-468**, 1045-1048 (2001).
11. K. Momma, F. Izumi, VESTA 3 for three-dimensional visualization of crystal, volumetric and morphology data. *J. Appl. Crystallogr.* **44**, 1272 (2011).

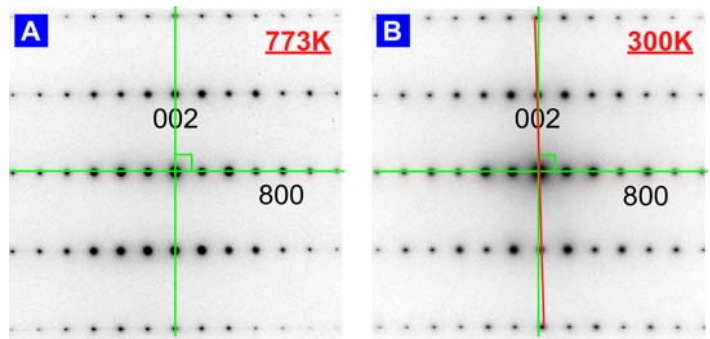

**Fig. S1.** Electron Diffraction patterns of BSO observed by TEM at (**A**) 793 K and (**B**) 298 K. The red line in panel **B** runs along the crystallographic *c*-direction. Crossed green lines, which bisect at right angles, are visual guides.

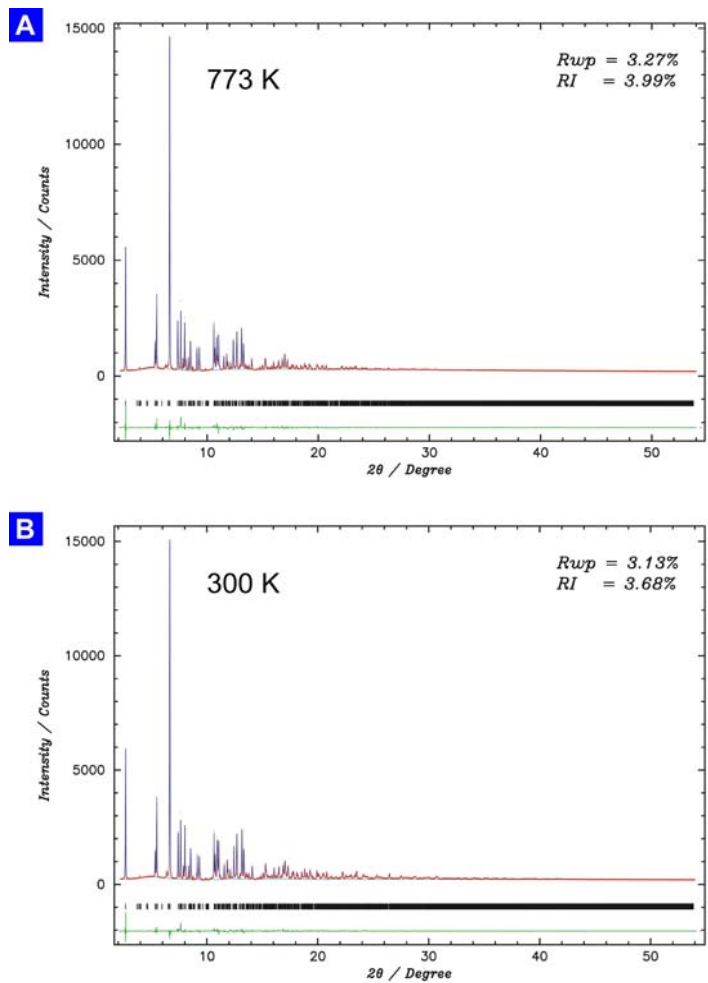

**Fig. S2.** X-ray powder diffraction patterns observed at (**A**) 773 K and (**B**) 300 K. Rietveld analyses assuming *Cmcm* and *Cc* converges with the values of $R_{wp}$ / $R_I$ of 3.27 % / 3.99 % and 3.13 % / 3.68 % for 773 K and 300 K, respectively. The calculated lines and residues are indicated in the panels by red and green lines, respectively.

| Cmcm | x | y | z | B | $U_{11}$ | $U_{22}$ | $U_{33}$ |
|---|---|---|---|---|---|---|---|
| Bi | 0.1647(1) | 0.2179(1) | 0.25 | | 0.0245(3) | 0.0219(3) | 0.0245(3) |
| Si | 0 | 0.6859(10) | 0.25 | | 0.0186(37) | 0.0220(45) | 0.0162(41) |
| O1 | 0 | 0.5 | 0.5 | | 0.0333(107) | 0.0404(101) | 0.0493(131) |
| O2 | 0.0918(4) | 0.1812(15) | 0.75 | 3.0485(2193) | | | |
| O3 | 0.2484(9) | 0.5 | 0.5 | 3.0485(2193) | | | |

| Cc | x | y | z | B | $U_{11}$ | $U_{22}$ | $U_{33}$ |
|---|---|---|---|---|---|---|---|
| Bi1 | 0.1696(1) | 0.2166(2) | 0.2609(7) | | 0.0082(2) | 0.0075(2) | 0.0074(3) |
| Bi2 | 0.8399(1) | 0.2195(2) | 0.2740(5) | | 0.0082(2) | 0.0075(2) | 0.0074(3) |
| Si | 0.4957(2) | 0.1720(6) | 0.2459(9) | | 0.0132(33) | 0.0098(34) | 0.0121(36) |
| O1 | 0.4990(24) | 0.0166(17) | 0.5112(49) | | 0.0216(86) | 0.0313(85) | 0.0313(85) |
| O2 | 0.5858(3) | 0.3309(22) | 0.2247(41) | 2.2564(4151) | | | |
| O3 | 0.4039(3) | 0.3233(23) | 0.2268(42) | 2.2564(4151) | | | |
| O4 | 0.2497(16) | 0.4942(98) | 0.4949(61) | 2.2564(4151) | | | |
| O5 | 0.2563(17) | 0.0065(125) | 0.0349(101) | 2.2564(4151) | | | |

**Table S1.** Structure parameters obtained by the Rietveld analyses for the *Cmcm* phase at 773 K (upper) and the *Cc* phase at 300 K (lower).

| Unstable Phonon Mode in *Cmcm* | Γ | Γ + shear distortion ($\beta \neq 90$) | $Y_1$ ($\omega_1$) | $Y_1$ + shear distortion ($\beta \neq 90$) | $Y_2$ ($\omega_2$) $|\omega_1| > |\omega_2|$ | $Y_2$ + shear distortion ($\beta \neq 90$) | S |
|---|---|---|---|---|---|---|---|
| Space Group | *Cmc2$_1$* | *Cc* | *Pnma* | *P2$_1$/c* | *Pbcn* | *P2c* | *P2$_1$/c* |
| Energy Decrease from *Cmcm* (meV) | -22.9 | -23.1 | -9.9 | -9.9 | -15.1 | -15.9 | -4.4 |

**Table S2.** Decrease in total energy from the paraelectric *Cmcm* structure by assuming lattice distortions, which correspond to the unstable phonon modes observed in the phonon dispersion for the paraelectric phase (Fig. 4A).

|  | *Cmcm* | | |
|---|---|---|---|
| *a* (Å) | 14.9217 | | |
| *b* (Å) | 5.4313 | | |
| *c* (Å) | 5.2376 | | |
|  |  |  |  |
| Bi | 0.1644 | 0.2222 | 1/4 |
| Si | 0 | 0.6828 | 1/4 |
| O1 | 0 | 1/2 | 1/2 |
| O2 | 0.0937 | 0.1724 | 3/4 |
| O3 | 0.2490 | 1/2 | 1/2 |

|  | *Cc* | | |
|---|---|---|---|
| *a* (Å) | 14.9803 | | |
| *b* (Å) | 5.4245 | | |
| *c* (Å) | 5.2582 | | |
| *β* (°) | 90.0136 | | |
|  |  |  |  |
| Bi1 | 0.1646 | 0.2232 | 0.2671 |
| Bi2 | 0.8354 | 0.2232 | 0.2672 |
| Si | 0.5000 | 0.1801 | 0.2261 |
| O1 | 0.5000 | 0.0304 | 0.4982 |
| O2 | 0.5928 | 0.3265 | 0.2022 |
| O3 | 0.4072 | 0.3266 | 0.2021 |
| O4 | 0.2494 | 0.4931 | 0.5019 |
| O5 | 0.2506 | 0.0069 | 0.0019 |

**Table S3.** Optimized lattice parameters of $Bi_2SiO_5$ obtained from DFT calculations.

| | |
|---|---|
| $\varepsilon(\infty)$ | $\begin{pmatrix} 4.58 & 0 & 0 \\ 0 & 5.31 & 0 \\ 0 & 0 & 5.32 \end{pmatrix}$ |
| $\varepsilon(0)$ | $\begin{pmatrix} 10.66 & 0 & 0 \\ 0 & 27.63 & 0 \\ 0 & 0 & 43.22 \end{pmatrix}$ |
| Bi | $\begin{pmatrix} 3.34 & 0.59 & 0 \\ 0.41 & 5.01 & 0 \\ -0.00 & 0.00 & 4.74 \end{pmatrix}$ |
| Si | $\begin{pmatrix} 4.75 & 0.00 & 0 \\ 0 & 3.33 & 0 \\ 0 & -0.00 & 3.24 \end{pmatrix}$ |
| O1 | $\begin{pmatrix} -2.25 & 0 & 0 \\ 0 & -2.33 & 1.04 \\ 0 & 1.03 & -2.71 \end{pmatrix}$ |
| O2 | $\begin{pmatrix} -2.67 & 0.95 & 0 \\ 0.90 & -2.52 & 0 \\ 0.00 & -0.00 & -2.18 \end{pmatrix}$ |
| O3 | $\begin{pmatrix} -1.93 & 0 & 0 \\ 0.00 & -3.00 & -0.03 \\ 0.00 & 0.17 & -2.81 \end{pmatrix}$ |

**Table S4.** Born effective charges and permittivity tensors of Bi$_2$SiO$_5$ in the *Cmcm* phase obtained from DFT calculations based on the linear response method.

| | |
|---|---|
| $\varepsilon(\infty)$ | $\begin{pmatrix} 4.58 & 0 & 0 \\ 0 & 5.21 & 0 \\ 0 & 0 & 5.26 \end{pmatrix}$ |
| $\varepsilon(0)$ | $\begin{pmatrix} 11.52 & 0 & -0.04 \\ 0 & 26.35 & 0 \\ -0.04 & 0 & 37.24 \end{pmatrix}$ |
| Bi1 | $\begin{pmatrix} 3.39 & 0.55 & -0.22 \\ 0.51 & 4.79 & 0.05 \\ -0.11 & 0.22 & 4.78 \end{pmatrix}$ |
| Bi2 | $\begin{pmatrix} 3.38 & -0.55 & 0.22 \\ -0.51 & 4.79 & 0.05 \\ 0.11 & 0.23 & 4.78 \end{pmatrix}$ |
| Si | $\begin{pmatrix} 4.73 & 0.00 & 0.00 \\ 0.00 & 3.32 & -0.30 \\ 0.00 & -0.31 & 3.27 \end{pmatrix}$ |
| O1 | $\begin{pmatrix} -2.23 & 0.00 & 0.00 \\ 0.00 & -2.28 & 0.96 \\ 0.00 & 1.00 & -2.76 \end{pmatrix}$ |
| O2 | $\begin{pmatrix} -2.69 & -0.91 & 0.38 \\ -0.84 & -2.38 & -0.06 \\ 0.23 & -0.18 & -2.19 \end{pmatrix}$ |
| O3 | $\begin{pmatrix} -2.69 & 0.91 & -0.38 \\ 0.84 & -2.39 & -0.06 \\ -0.23 & -0.18 & -2.19 \end{pmatrix}$ |
| O4 | $\begin{pmatrix} -1.95 & -0.08 & -0.10 \\ -0.22 & -2.93 & -0.12 \\ -0.13 & 0.18 & -2.84 \end{pmatrix}$ |
| O5 | $\begin{pmatrix} -1.96 & -0.08 & 0.10 \\ -0.23 & -2.92 & 0.12 \\ 0.14 & -0.18 & -2.84 \end{pmatrix}$ |

**Table S5.** Born effective charges and permittivity tensors of $Bi_2SiO_5$ in the $Cc$ phase obtained from DFT calculations based on the linear response method.

**Movie S1.** An animation for the soft mode. It should be noted that the absolute value of amplitude is arbitrary in this animation because a normalized eigenvector is obtained from a dynamical matrix. Displacement ratios among atoms, however, are physically meaningful.